\documentclass[twocolumn, secnumarabic,amssymb, aps]{revtex4-1}

\usepackage[columnwise]{lineno}

\usepackage[utf8]{inputenc}
\usepackage{CJKutf8}
\usepackage{amsmath,amssymb}
\usepackage{type1cm}
\usepackage{mathrsfs}
\usepackage{mathtools}
\usepackage{url}
\usepackage{graphicx}
\usepackage{color}
\usepackage{booktabs}
\newcommand\HL[1]{{\color{black}#1}}
\newcommand\HLL[1]{{\color{black}#1}}
\newcommand\HLPOF[1]{{\color{black}#1}}
\usepackage{bm}
\newcommand{\vect}[1]{\boldsymbol{\mathbf{#1}}}
\def\vec#1{\vect{#1}}

\begin{document}
\title{Theoretical framework bridging classical and quantum mechanics for the dynamics of cryogenic liquid helium-4 using smoothed-particle hydrodynamics}

\author{Satori Tsuzuki~(\begin{CJK}{UTF8}{min}都築怜理\end{CJK})}
\email[Email:~]{tsuzukisatori@g.ecc.u-tokyo.ac.jp\\ \url{https://www.satoritsuzuki.org/} }
\affiliation{Research Center for Advanced Science and Technology, University of Tokyo, 4-6-1, Komaba, Meguro-ku, Tokyo 153-8904, Japan}
\begin{abstract}
Our recent study suggested that a fully classical mechanical approximation of the two-fluid model of superfluid helium-4 based on smoothed-particle hydrodynamics (SPH) is equivalent to solving a many-body quantum mechanical equation under specific conditions. This study further verifies the existence of this equivalence. First, we derived the SPH form of the motion equation for the superfluid component of the two-fluid model, i.e., the motion equation driven by the chemical potential gradient obtained using the Gibbs--Duhem equation. We then derived the SPH form of the motion equation for condensates based on the Gross--Pitaevskii theory, i.e., the motion equation driven by the chemical potential gradient obtained from the Schr${\rm \ddot{o}}$dinger equation of interacting bosons. Following this, we compared the two discretized equations. Consequently, we discovered that a condition maintaining zero internal energy for each fluid particle ensures the equivalence of the equations when the quantum pressure is negligible. Moreover, their equivalence holds even when the quantum pressure is nonnegligible if the quantum pressure gradient force equals the mutual friction force. A zero internal energy indicates the thermodynamic ground state, which includes an elementary excitation state. Therefore, the condition can be sufficiently satisfied when the velocities of fluid particles do not exceed the Landau critical velocity, which is not a stringent condition for simulations with a characteristic velocity of a few $\rm cm\cdot s^{-1}$ in a laboratory system. Based on the above, we performed a simulation of rotating liquid helium-4 and succeeded in generating a vortex lattice with quantized circulation, known as a quantum lattice.
\end{abstract}
\maketitle

\begin{figure*}[t]
\vspace{+0.5cm}
\centerline{\includegraphics[scale=0.53]{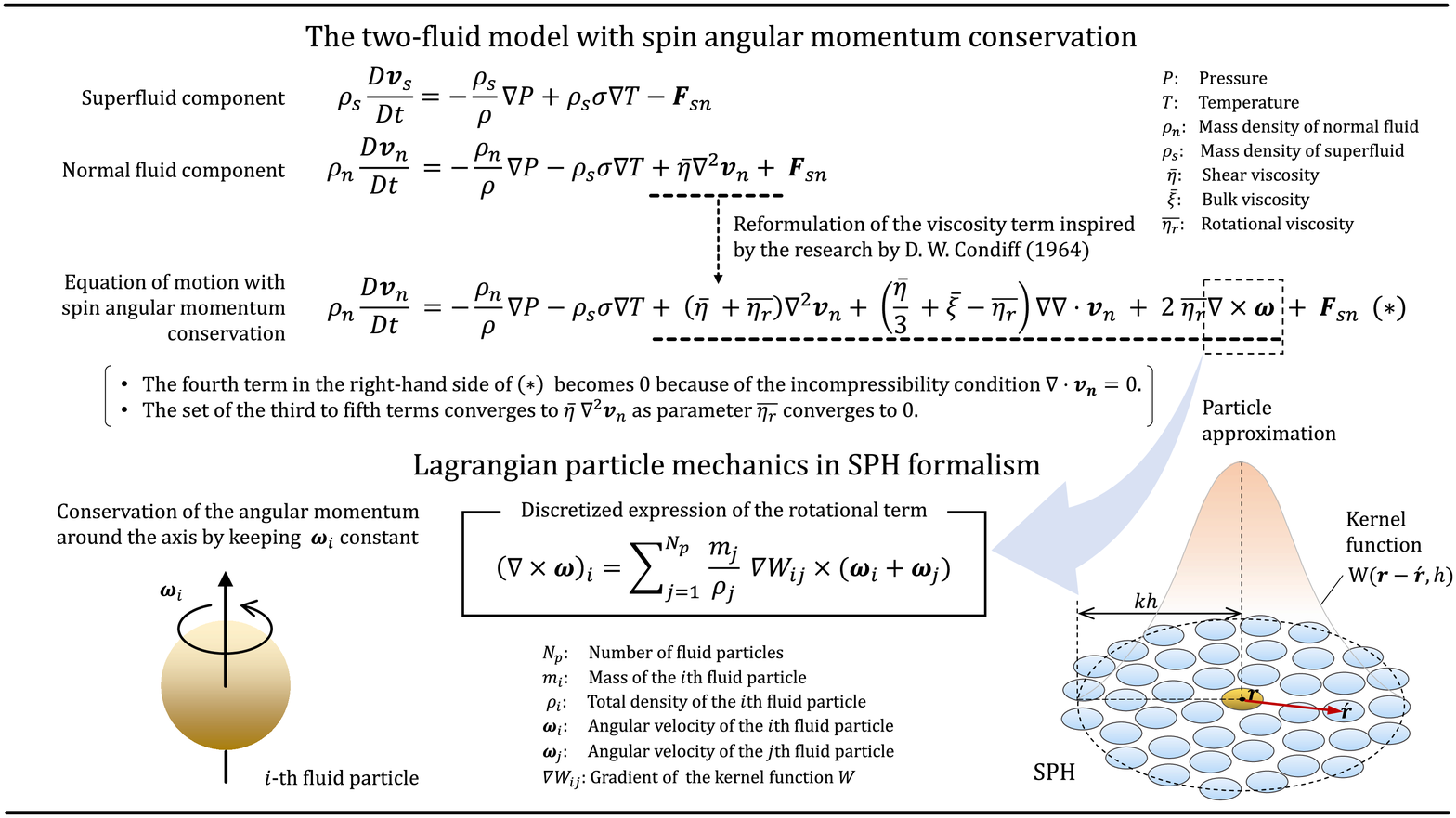}}
\caption{Schematic of the two-fluid model with angular momentum conservation.}
\label{fig:OverviewPrevOurWork}
\end{figure*}

\section{Introduction}
Since the beginning of the 20th century, the significant decrease in the viscosity of liquid helium-4 in cryogenic regions (approximately 2.1 K) has attracted the attention of researchers in the field of low-temperature physics. 
\HLPOF{The viscosity loss of liquid helium-4 as a continuum corresponds microscopically to the loss of molecular viscosity, which results from the van der Waals forces between helium atoms. This loss is even more microscopically attributed to the innate nature of atoms, i.e., the change in the energy states of helium atoms manifested in the cryogenic temperature range.}
Notably, film flow is a phenomenon observed when liquid helium-4 creeps out of its container as it loses its viscosity and becomes a superfluid. \HL{The} fountain effect, caused by a superleak of helium atoms passing through a porous medium, such as a plaster, is \HL{also} well known. 
\HLPOF{More precisely, the ``fountain effect'' refers to the thermodynamic effect observed in a vessel immersed in a reservoir, which is equipped with a heater, an opening at the top, and a capillary filter at the bottom. This arrangement allows the flow of a superfluid into the heater and causes the normal fluid generated by the heater to flow out through the opening. Because the viscosity loss properties of helium atoms in this temperature range enable them to penetrate such capillary filters, this penetration phenomenon is referred to as a ``superleak.''} 

\HLPOF{Understanding the dynamics governing the behavior of} superfluid helium-4 is not only an outstanding academic achievement \HLPOF{in low-temperature physics} but is also expected to clarify various related phenomena owing to the similarities with the dynamics of the phase transition of helium-4. Some related examples are liquid crystals~\cite{Kai_1975, Halinen2000, PhysRevB.94.180501, doi:10.1126/science.251.4999.1336} and superconductivity~\cite{tilley2019superfluidity, parker2008vortices} in condensed matter physics, Hawking radiation~\cite{HAWKING1974, Hawking1975, PhysRevD.98.124043, PhysRevA.68.053613} and neutron star crusts~\cite{10.1143/PTPS.186.9} in astrophysics, and turbulence~\cite{PhysRevLett.129.025301, doi:10.1063/1.4898666, PhysRevX.12.011031}. 
\HLPOF{In addition to the value of such a scientific study, realizing robust and safe control of superfluid helium-4 in the bulk state by predicting its complex behavior under various conditions via large-scale fluid simulations is of paramount importance in the development of cryogenic cooling systems for X-ray observation satellites~\cite{YOSHIDA201827, EZOE2020103016} and space telescopes~\cite{Woods2020, 10.1117/1.JATIS.7.1.011008} for high-energy astronomy; these simulations may be based on classical fluid dynamics while being capable of estimating the effects of macroscopic quantum phenomena.} Therefore, a direct numerical analysis of superfluid helium-4 in the bulk state on scales ranging from centimeters to meters can be expected to facilitate the development of safe and robust cryogenic cooling systems and significantly enhance the engineering applicability of quantum liquids.

However, no method for simulating the behavior of such large-scale superfluid helium-4 has yet been established. Generally, methods that can simulate superfluid helium-4 can be categorized into \HLPOF{four} major types. The first involves treating superfluid helium as a quantum many-body system with a Bose--Einstein condensate~\cite{Bose1924, PhysRevLett.94.050402, PhysRevLett.101.010402, PhysRevA.81.025604} and directly solving the Schr${\rm \ddot{o}}$dinger equation. The second is the vortex filament model (VFM), which solves the dynamics of superfluid fields using a Lagrangian approach~\cite{Idowu2001, doi:10.1063/1.4828892, PhysRevLett.120.155301, doi:10.1063/1.5091567, PhysRevLett.124.155301}. \HLPOF{The third method is the density functional theory (DFT), which is also a well-established method for obtaining the density states~\cite{PhysRev.136.B864, doi:10.1073/pnas.76.12.6062, PhysRevLett.59.2360}. Examples of the applications of the DFT to liquid helium-4 include studies on nanodroplets~\cite{Barranco2006, PhysRevB.91.100503} and electron bubble states~\cite{doi:10.1063/1.2745297, doi:10.1063/1.3544216}.}
These \HLPOF{three} methods are quantum mechanical models that describe phenomena on the order of nanometers to micrometers. However, using them in the simulations of macroscopic cryogenic liquid helium-4 would require an excess of tens of billions of atomic-sized analytic particles, which is unrealistic even with the latest supercomputers. 
In contrast, the \HLPOF{fourth} method is a phenomenological model proposed by Landau and Tisza~\cite{PhysRev.60.356, TISZA1938}, known as the two-fluid model. It describes liquid helium-4 based on two components: a superfluid component that is an inviscid incompressible fluid and a normal fluid component that follows the incompressible Navier--Stokes equation; the model describes the system as a mixture or superposition of these components. However, this is a classical fluid approximation, and it cannot reproduce macroscopic quantum phenomena, such as a vortex lattice in rotating liquid helium-4. 

Recently, an improved two-fluid model, including the conservation of angular momentum, was developed in our previous study~\cite{Tsuzuki_2021, doi:10.1063/5.0060605}. This model has attracted \HLPOF{attention} as a new approximation model that balances classical and quantum mechanical descriptions to address the drawbacks of existing approaches. Our previous study~\cite{Tsuzuki_2021} reformulated the motion equation of the normal fluid component of the two-fluid model to include a term accounting for the angular momentum conservation of particles around their axes. Let us refer to this rotational angular momentum as the ``spin angular momentum'' analogous to the corresponding term in quantum mechanics. As will be discussed later, our previous study~\cite{Tsuzuki_2021} treated the fluid forces of both components as a classical mechanical approximation and discretized the reformulated two-fluid model using smoothed-particle hydrodynamics~(SPH), which is a well-established Lagrangian particle approximation model for flow problems originally developed in astrophysics~\cite{gingold1977smoothed}. Interestingly, we observed the emergence of parallel spinning vortices presenting rigid-body rotation in the numerical simulation of rotating liquid helium-4. In addition, our subsequent study~\cite{doi:10.1063/5.0060605} incorporated a vortex dynamics model into \HL{the reformulated} two-fluid model, successfully capturing the phenomenon of vortex lattices via numerical simulations.\cite{doi:10.1063/5.0060605} also reported that the number of vortices generated in a system of rotating cylinders agreed with the theoretical solution based on Feynman's rule~\cite{feynman1955progress, VANSCIVER2009247}. However, it still included one free parameter that had to be optimized to determine the intensity of the angular velocity. A video of the simulation in \cite{doi:10.1063/5.0060605} is available at {\small \url{https://arxiv.org/src/2105.03177v3/anc}}. 
These results indicate that a vortex lattice in rotating liquid helium-4 can be reproduced even by solving the two-fluid model based on a fully classical mechanical approximation that includes the fluid forces of both components. This is applicable under the condition that the viscosity is rederived to conserve the rotational angular momentum using the SPH and the vortex dynamics \HLPOF{is incorporated into the system}. In particular, a vortex lattice in rotating liquid helium-4 is \HL{deemed} a purely quantum mechanical phenomenon. Thus, the results reported in ~\cite{doi:10.1063/5.0060605} have gained significant scientific interest because they directly challenge the decade-long preconception that a fully classical fluid dynamics approach cannot reproduce quantum mechanical phenomena, such as vortex lattices in liquid helium-4.

Figure~\ref{fig:OverviewPrevOurWork} presents a schematic of the improved two-fluid model developed in our previous study ~\cite{Tsuzuki_2021, doi:10.1063/5.0060605}. \HL{The} involved method has two essential features. The first is the reformulation of the viscosity term in the normal fluid component of the two-fluid model. Although the ordinary two-fluid model can describe the macroscopic dynamics of cryogenic liquid helium-4, angular momentum conservation, which is the essence of quantum mechanics, has not been explicitly formulated. In this regard, in a study on numerical simulations of mesoscale flows considering microfluidics, the Navier--Stokes equation with angular momentum conservation for the rotational motion of molecules constituting a polar fluid was formulated by Condiff and Dahler~\cite{doi:10.1063/1.1711295}. Moreover, the conservation of angular velocities of fluid particles \HL{was} explicitly rederived in the Lagrangian form. In addition, in recent studies, a method~\cite{MULLER2015301} has been developed to discretize this rederived Navier--Stokes equation using smoothed-dissipative particle dynamics~\cite{PhysRevE.67.026705} for the simulation of mesoscale flows, such as a blood flow, with heat dissipation~\cite{PhysRevE.95.063314, doi:10.1073/pnas.1608074113}. These studies led us to believe that applying the same technique to the discretization of the motion equation for the normal fluid component may enable angular momentum conservation in the two-fluid model, thereby making it feasible to reproduce the macroscopic quantum phenomena of liquid helium-4. The second feature is \HL{the} finite particle approximation using Lagrangian particle mechanics, which is well suited for these types of scenarios\HL{,} i.e., scenarios regarding fluid particles as coarse-grained helium atoms. It also facilitates spin angular momentum conservation of the fluid particles by maintaining the angular velocity constant for each spherical fluid particle, as illustrated in the lower left side of Fig.~\ref{fig:OverviewPrevOurWork}. In our previous studies~\cite{Tsuzuki_2021, doi:10.1063/5.0060605}, the SPH was employed \HL{to this end}.

\HL{In addition}, greater emphasis should be placed on our classical approximation detailed in ~\cite{Tsuzuki_2021, doi:10.1063/5.0060605}, which includes the fluid forces of both components and solves their motion equations in a fully coupled manner, such as a multiphase flow in classical fluid dynamics. We will further describe this point. The original two-fluid model proposed by Landau treats the aforementioned two components independently; however, the existence of mutual friction forces was proposed by Gorter and Mellink~\cite{GORTER1949285}. \HL{Following this}, the coupling of the two components was gradually confirmed. In a recent related study on liquid helium-4, the superfluid component was solved using the VFM and coupled with the Navier--Stokes equation~\cite{Idowu2001, PhysRevLett.120.155301, doi:10.1063/1.5091567, PhysRevLett.124.155301}. In our \HL{previous studies}, we directly coupled the Navier--Stokes equation not with the VFM but with the motion equation of the superfluid component in the classical two-fluid model, i.e., we employed the motion equation of an inviscid incompressible flow. \HL{For quantum mechanical corrections}, we conserved the spin angular momentum to ensure close resemblance between the fluid system and a quantum mechanical system. In a broader sense, among these existing coupling approaches, our approach can be classified as one with a coarse-grained model.

\HL{Importantly}, we observed the \HL{emergence} of multiple \HL{spinning} vortices forming a \HL{rigid-body} lattice, despite solving the two-fluid model based on a fully classical approximation incorporating the fluid forces of both components. \HL{Therefore, regardless of the above}, the following hypothesis can be suggested: discretizing the two-fluid model using particles in the SPH formalism can extract the nature of a multiparticle interacting system, and a fully classical mechanical approximation \HL{may be} equivalent to solving a many-body quantum mechanical equation under specific conditions. This hypothesis is \HL{valid only} for large-scale problems because the fully classical approximation disregards several laws of quantum mechanics. In \HL{our previous studies}, the \HL{SPH discretization} of the two-fluid model \HL{with angular momentum conservation} and its application to large-scale problems of rotating liquid helium-4 made it possible to observe the phenomena of vortex lattices. However, no study has provided theoretical evidence supporting the equivalence of the microscopic equation of motion of a many-body quantum system and the phenomenological motion equation of \HL{the superfluid component of the two-fluid model} in the SPH formalism.

To address this gap, this study aims to demonstrate the existence of this equivalence. First, we derive the SPH form of the motion equation of the superfluid component of the two-fluid model, i.e., the motion equation driven by the gradient of the chemical potential \HL{obtained using} the Gibbs--Duhem equation. Then, we derive the SPH form of the motion equation for condensates from the Gross--Pitaevskii (GP) theory, i.e., the motion equation driven by the gradient of the chemical potential obtained from the Schr${\rm \ddot{o}}$dinger equation of interacting bosons. We then compare these two discretized motion equations in their SPH forms derived independently from microscopic and macroscopic perspectives to identify a condition wherein both become equivalent. 
Notably, we report that maintaining the internal energy zero for each fluid particle ensures the above equivalence. We also discuss the implications of this finding in terms of real-world cases and detail its incorporation into liquid helium-4 simulations as a stepping stone toward accurate future numerical reproduction of macroscopic quantum phenomena, such as film flows and fountain phenomena. 


\section{Brief review of existing theories}
\subsection{Gross--Pitaevskii theory} \label{seq:GPtheory}
The time-dependent many-body Schr${\rm \ddot{o}}$dinger equation and its quantum Hamiltonian $H$ are as follows~\cite{Salasnich2017BSIUA}:
\begin{eqnarray}
i\hbar\frac{\partial }{\partial t} \psi(\vec{r}_{1}, \vec{r}_{2}, \cdots, \vec{r}_{N}, t) = H \psi(\vec{r}_{1}, \vec{r}_{2}, \cdots, \vec{r}_{N}, t),~~~~~\\
H = - \frac{\hbar^2}{2 m_q}\sum_{i=1}^{N} \nabla_{i}^2 + \sum_{i=1}^{N} U(\vec{r}_{i}, t) + \frac{1}{2} \sum_{i \ne j}^{N} V(\vec{r}_{i} - \vec{r}_{j}) \label{eq: Hamiltonian},~~~~~
\end{eqnarray}
where $\psi(\vec{r}_{1}, \vec{r}_{2}, \cdots, \vec{r}_{N}, t)$ represents a many-body wavefunction, $N$ denotes the number of particles, $m_q$ denotes the mass of a particle, $\hbar$ denotes the reduced Planck's constant, $U(\vec{r}_{i}, t)$ represents the external potential, and $V(\vec{r}_{i} - \vec{r}_{j})$ represents the mutual interaction potential between the $i$th and $j$th particles. 

For $N$ identical interacting bosons, the wavefunction $\psi(\vec{r}_{1} \vec{r}_{2}, \cdots, \vec{r}_{N}, t)$ is symmetric with respect to the exchange between the $i$th and $j$th particles, as follows:
\begin{eqnarray}
\psi(\vec{r}_{1}, \cdots, \vec{r}_{i}, \cdots, \vec{r}_{j}, \cdots, \vec{r}_{N}, t) \nonumber~~~~~\\
= \psi(\vec{r}_{1}, \cdots, \vec{r}_{j}, \cdots, \vec{r}_{i}, \cdots, \vec{r}_{N}, t). 
\end{eqnarray}
The Bose--Einstein condensate~\cite{Bose1924} assumes that all particles occupy the same single-particle state. Thus, the many-body wavefunction $\psi(\vec{r}_{1}, \vec{r}_{2}, \cdots, \vec{r}_{N}, t)$ can be decomposed into a product of single-particle wavefunctions $\phi(\vec{r}_{i}, t)$, as follows~\cite{Salasnich2017MBS, hartree_1928, pethick2008bose}: 
\begin{eqnarray}
\psi(\vec{r}_{1}, \vec{r}_{2}, \cdots, \vec{r}_{N}, t)~& \coloneqq &~\prod_{i=1}^{N} \phi(\vec{r}_{i}, t), \label{eq:hartreeprod}
\end{eqnarray}
where $\phi(\vec{r}_{i}, t)$ satisfies the normalization condition as follows:
\begin{eqnarray}
	\int |\phi(\vec{r}, t)|^2 d\vec{r} &=& 1. \label{eq:phiNormalCond}
\end{eqnarray}

The condensate wavefunction is described as follows:
\begin{eqnarray}
\Psi(\vec{r}) \coloneqq \sqrt{N}\phi(\vec{r}). \label{eq:DefPsi}
\end{eqnarray}
Based on the definition given in Eq.~(\ref{eq:DefPsi}) and the normalization condition given in Eq.~(\ref{eq:phiNormalCond}), we can obtain $\int |\Psi|^2 d\vec{r} = N$. 
Therefore, the condensate wavefunction $\Psi$ can be expressed as a function of the condensate density $n(\vec{r}, t)$, as follows~\cite{Barenghi2016}:
\begin{eqnarray}
\Psi(\vec{r}, t) = \sqrt{n(\vec{r}, t)} {\rm e}^{i \theta(\vec{r}, t)}, \label{eq:WaveFuncPsi}
\end{eqnarray}
where $i$ denotes an imaginary unit, and $\theta(\vec{r}, t)$ represents the phase of the condensate wavefunction. 

Substituting the Lagrangian function of the Hamiltonian given in Eq.~(\ref{eq: Hamiltonian}) into the Euler--Lagrangian equation to satisfy the least-action principle yields the following time-dependent GP equation for the condensed matter wavefunction~\cite{mishmash2008quantum, kunimi2014}:
\begin{eqnarray}
i \hbar\frac{\partial}{\partial t} \Psi(\vec{r}, t) = &-&\frac{\hbar^2}{2 m_q} \nabla^2 \Psi(\vec{r},t) \nonumber \\
&+& U(\vec{r}, t)\Psi(\vec{r},t) \nonumber \\
&+& \int V(\vec{r} - \vec{\acute{r}}) |\Psi(\vec{\acute{r}},t)|^2 d\vec{\acute{r}}. \label{eq:TimeDepGP}
\end{eqnarray}
By substituting Eq.~(\ref{eq:WaveFuncPsi}) into Eq.~(\ref{eq:TimeDepGP}) and comparing the real and imaginary parts of both sides of Eq.~(\ref{eq:TimeDepGP}), we can obtain the following series of equations for the condensate density $n(\vec{r},t)$~\cite{kunimi2014, Barenghi2016}:
\begin{eqnarray}
n(\vec{r}, t) = &-& \nabla \cdot \Bigl[ n(\vec{r}, t) \frac{\hbar}{m_q} \nabla \theta(\vec{r}, t) \Bigr], \\
\hbar\frac{\partial}{\partial t} \theta(\vec{r},t) = &-&\frac{\hbar^2}{2 m_q} \bigl[ \nabla \theta(\vec{r}, t) \bigr]^2 \nonumber \\
&-& U(\vec{r}, t) \nonumber \\ 
&-& \int V(\vec{r} - \vec{\acute{r}}) n(\vec{\acute{r}}, t) d\vec{\acute{r}} \nonumber \\ 
&+& \frac{\hbar^2}{2 m_q} \frac{\nabla^2 \sqrt{n(\vec{r},t)}}{\sqrt{n(\vec{r}, t)}}, \label{eq:MomentumEqNThetaiVer}
\end{eqnarray}
where we can deduce the relationship between the velocity of the superfluid component and the phase of the condensate wavefunction as follows:
\begin{eqnarray}
\vec{v}_{s} (\vec{r}, t) &\coloneqq& \frac{\hbar}{m_q} \nabla \theta(\vec{r}, t). \label{eq:SupeVelDefPhase}
\end{eqnarray}

\subsection{Smoothed-particle hydrodynamics} \label{sec:explainsph}
In SPH, a discrete physical quantity $\varphi$ is represented as a continuous quantity using the Dirac delta function, $\delta$, which is further approximated using a distribution function $W$, known as the kernel function~\cite{monaghan1992smoothed}, as follows:
\begin{eqnarray}
\varphi(\vec{r}) &=& \int_{\Omega} \varphi(\vec{r})\delta(\vec{r} - \vec{\acute{r}}) d\vec{\acute{r}} \label{eq:defDiracPhi}\\
&\simeq& \int_{\Omega} \varphi(\vec{r})W(\vec{r} - \vec{\acute{r}}, h) d\vec{\acute{r}}, \label{eq:SPHapprox}
\end{eqnarray}
where kernel function $W$ is selected to satisfy the following conditions:
\begin{eqnarray}
\lim_{h \to 0} W(\vec{r} - \vec{\acute{r}}, h) &=& \delta (\vec{r} - \vec{\acute{r}}), \label{eq:Wcond:conv} \\
W(\vec{r}) &=& W(-\vec{r}), \label{eq:Wcond:symm} \\	
\int W d\vec{r} &=& 1. \label{eq:Wcond:norm}
\end{eqnarray}
A straightforward example of $W$ satisfying the given conditions in Eqs.~(\ref{eq:Wcond:conv})--~(\ref{eq:Wcond:norm}) is the Gaussian kernel~\cite{gingold1977smoothed} expressed as $W(\vec{r} - \vec{\acute{r}}) = C/h^d {\rm exp}[-|\vec{r} - \vec{\acute{r}}|^2 /h^2]$, where $C$ denotes a normalization constant, $h$ denotes the kernel radius, and $d$ represents the dimension. However, in the simulations of incompressible flows, a polynomial function~\cite{desbrun1996smoothed, muller2003particle} that converges to zero at a distance of $kh$ is selected as the kernel function because it satisfies the normalization condition given in Eq.~(\ref{eq:Wcond:norm}) by simply integrating over a distance $kh$ without having to integrate over the entire domain. In this study, a cubic spline kernel function~\cite{1985AA149135M} is used in the numerical experiments described in Section~\ref{seq:apptopra}. 

The gradient of $\varphi$ is expressed as follows:
\begin{eqnarray}
\nabla \varphi(\vec{r}) &\coloneqq& \int_{\Omega} \nabla \varphi(\vec{r})W(\vec{r} - \vec{\acute{r}}, h) d\vec{\acute{r}}, \label{eq:SPHgradorg} \\ 
&\simeq& \int_{\Omega} \varphi(\vec{r})\nabla W(\vec{r} - \vec{\acute{r}}, h) d\vec{\acute{r}}. \label{eq:SPHgrad}
\end{eqnarray}
Here, Eq.~(\ref{eq:SPHgrad}) is derived from Eq.~(\ref{eq:SPHgradorg}) using Gauss's divergence theorem, after representing Eq.~(\ref{eq:SPHgradorg}) in the form of integration by parts~\cite{monaghan1992smoothed}.

Let us consider a case wherein the integral domain $\Omega$ is divided into $N_p$ small volumes $\Delta V_i$ ($i=1, 2, \cdots, N_p$). Based on the summation approximation, Eqs.~(\ref{eq:SPHapprox}) and ~(\ref{eq:SPHgrad}) can be expressed in discrete forms using mass $m_i$, density $\rho_i$, and small volume $\Delta V_i$ and their relationship $\Delta V_i = m_i/\rho_i$, as follows:
\begin{eqnarray}
\varphi(\vec{r}_i) &=& \sum_{j=1}^{N_p} \varphi(\vec{r}_j) \frac{m_j}{\rho_j}W_{ij}, \label{eq:SPHapproxdisc} \\
\nabla \varphi(\vec{r}_i) &=& \sum_{j=1}^{N_p} \varphi(\vec{r}_j) \frac{m_j}{\rho_j} \nabla W_{ij}, \label{eq:SPHgraddisc}
\end{eqnarray}
where $W_{ij} = W(|\vec{r}_{i}-\vec{r}_{j}|, h)$ and $\nabla W_{ij}$ denote the gradient of $W_{ij}$. 
Note that the remainder of our analysis only requires the expression of the gradient of $\varphi(\vec{r}_i)$ in Eq.~(\ref{eq:SPHgraddisc}).
Additional details on improved SPH operators for the gradient, basic and improved SPH operators for rotation and divergence, and several numerical techniques necessary for stable SPH simulations can be found elsewhere~\cite{doi:10.1063/5.0060605, MULLER2015301, monaghan1992smoothed, doi:10.1063/1.5068697}.
\begin{figure*}[t]
\centerline{\includegraphics[scale=0.64]{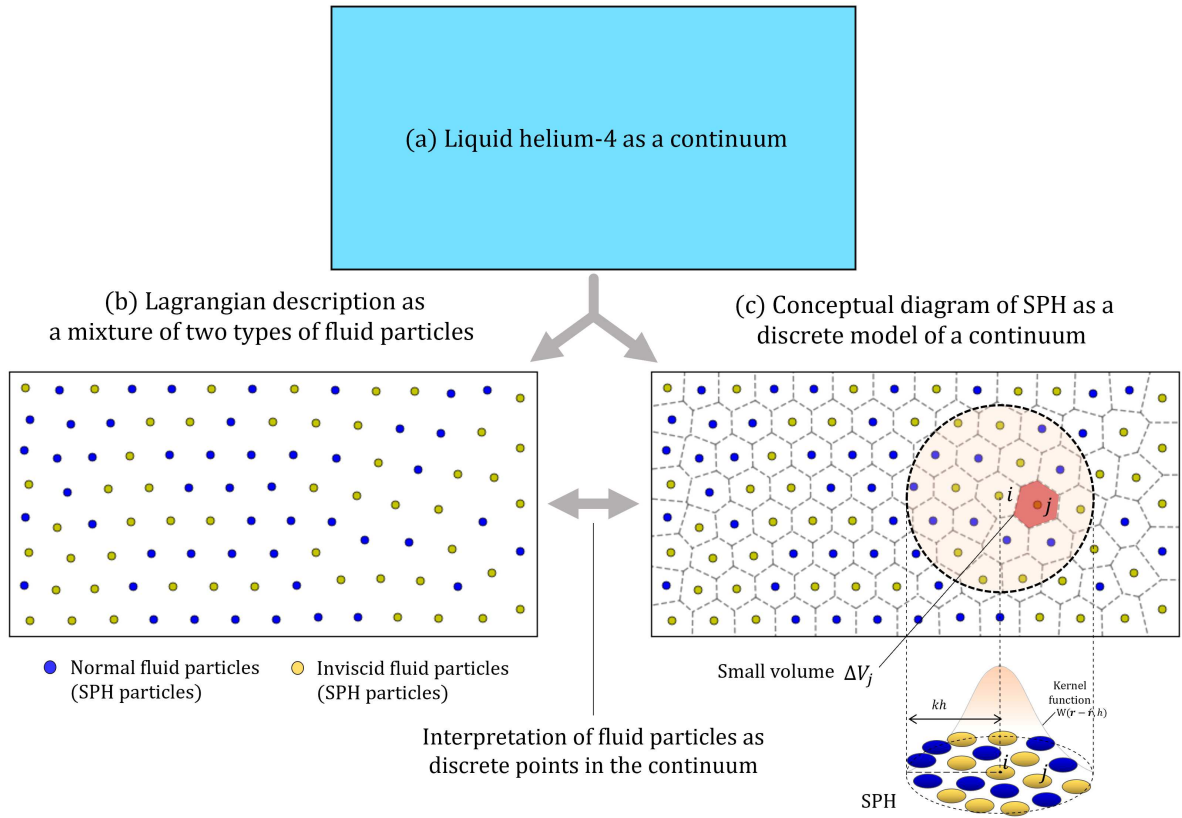}}
\caption{Schematics of SPH discretization of liquid helium-4: (a) liquid helium-4 as a continuum, (b) Lagrangian description as a mixture of two types of classical fluid particles: normal and inviscid fluid particles, and (c) conceptual diagram of SPH as a discrete model of continuum}
\label{fig:Figure-ConceptOfSPHForAContinuum}
\end{figure*}

\HLPOF{Before proceeding, we will attempt to clarify our concept of the SPH discretization of liquid helium-4 using Fig.~\ref{fig:Figure-ConceptOfSPHForAContinuum} and explain some of the terms referred in the subsequent sections. First, as already mentioned, we apply the classical Lagrangian picture, as depicted in Fig.~\ref{fig:Figure-ConceptOfSPHForAContinuum}(b), to the helium-4 continuum, which is presented in Fig.~\ref{fig:Figure-ConceptOfSPHForAContinuum}(a). Specifically, we assume a scenario wherein two types of classical fluid particles are mixed: normal fluid particles following the Navier--Stokes equation and inviscid fluid particles following the momentum equation of a classical inviscid flow. Importantly, when discretizing a continuum using SPH, fluid particles are assumed to correspond to the discretization points of the continuum. Fig.~\ref{fig:Figure-ConceptOfSPHForAContinuum}(c) illustrates an example of the application of the Voronoi tessellation, which is commonly used for the spatial discretization of classical incompressible fluids, to a domain of the helium-4 continuum. The gray dashed lines represent the boundaries of the Voronoi diagram. The black circle represents the neighboring area of the $i$th particle or the discretization point in the continuum. The bottom part of Fig.~\ref{fig:Figure-ConceptOfSPHForAContinuum}(c) depicts the particle distribution within the corresponding kernel function of the $i$th particle in the SPH calculation.}

\HLPOF{The most important point to note is that all fluid particles in this calculation are classical fluid particles (in fact, if these fluid particles are to obey quantum mechanics, they cannot be unambiguously labeled, as shown in Figs.~\ref{fig:Figure-ConceptOfSPHForAContinuum}(b) and (c), because identical particles cannot be distinguished in quantum mechanics). Thus, in this paper, ``normal fluid particles'' or ``superfluid particles'' refer to classical fluid particles unless otherwise stated. It should be emphasized that ``superfluid particles'' refer to classical fluid particles that obey the momentum equation of an inviscid flow. Moreover, note that only in Section~\ref{seq:GPtheory}, the term ``particle'' implies an atom; otherwise, it implies a classical fluid particle. As mentioned, the two types of classical fluid particles (normal and inviscid) denote the discretization points of the continuum, as depicted in Fig.~\ref{fig:Figure-ConceptOfSPHForAContinuum}(c). Specifically, they are virtual particles. This concept is based on several recent mathematical studies demonstrating that SPH is a generalized particle method, wherein each fluid particle does not serve as a physically significant particle but simply acts as a discretization point in a Voronoi cell finite element~\cite{imoto2019convergence, Imoto2020, IMOTO2022115012}. In our studies, we adopted this concept and postulated that these two types of virtual particles have no physical significance other than as discretization points, or fragments, of a continuous two-phase flow. Nevertheless, future research may provide some quantum physical interpretation of these virtual particles.}

\HLPOF{Because the two types of fluid particles maintain their volumes, both contribute to the macroscopic fluid phenomena uniquely observed as a whole. In summary, our model adopts a type of one-fluid system, wherein the fluid phenomena of the entire system are focused upon, whereas the motions of individual virtual particles have no physical significance from a microscopic viewpoint. The concept of a one-fluid system for liquid helium-4 is not very prominent; however, it has a long history~\cite{MONGIOVI20181}. Previous studies reviewed in~\cite{MONGIOVI20181} described the dynamics of liquid helium-4 from a microscopic viewpoint, primarily based on quantum mechanics. Our studies differ from these studies in that we attempted to describe identical dynamics in terms of the finite particle approximation of a continuum. Importantly, the concept of our virtual fluid particles is different from the two-component concept of the ordinary two-fluid model. This discussion highlights that the motions of individual virtual fluid particles do not represent the behavior of a normal fluid or superfluid component of the ordinary two-fluid model. We reemphasize the following. Our particle approximation is based on the concept of a generalized particle method wherein a fluid particle only serves as a discretization point in a continuum. Specifically, all fluid particles are classical fluid particles that have no physical significance from a microscopic viewpoint.} 

\HLPOF{The idea indicating that superfluid particles retain their volumes may be innovative. However, experiments have demonstrated that liquid helium-4 condensates produce only approximately 13\% of the total atoms even at approximately absolute zero~\cite{PhysRevLett.49.279}. Therefore, we do not observe a massive loss of liquid helium volume in the cryogenic range. Our model assuming a constant volume of inviscid fluid particles is consistent with these facts and has a certain validity from a phenomenological standpoint. Therefore, only the fluid phenomena of the system as a whole have physical significance. Because all particles are classical fluid particles, we refer to our approach as a ``fully classical mechanical approximation.'' By contrast, in the conventional two-fluid model, the macroscopic spatial domain is occupied by the normal fluid components, which microscopically interact with quantum vortices~\cite{doi:10.1063/1.4828892, doi:10.1063/1.4984913, PhysRevLett.124.155301}. Therefore, the computational results of fluid simulations based on our model, which employs a one-fluid system, must be compared with those of normal fluid component simulations of the two-fluid model based on conventional methods, which employ a two-fluid system. This is important when comparing the simulation results of this study with those of previous studies cited in Section~\ref{seq:apptopra}.} 

\HLPOF{In addition, the mutual friction forces are added to the system of equations, as depicted in Fig.~\ref{fig:OverviewPrevOurWork}, in our actual calculations. However, in the simulations based on our model, the magnitudes of the mutual friction forces are sufficiently smaller compared to those of the fluid forces, such as pressure gradient forces, when using the microscopic models of mutual friction forces reported in~\cite{NEMIROVSKII201385}. Nevertheless, we cannot deny the possibility that these terms may affect the system as disturbances in certain scenarios, even in our scheme, which is based on classical fluid mechanics. Furthermore, considering the future extension of our model to microscopic phenomena in the regime governed by quantum mechanics, retaining these terms at this stage may be useful for formal consistency with the general two-fluid model, wherein a mutual friction term is introduced. Therefore, we retained these mutual friction force terms in the SPH model in our series of studies.}

\HLPOF{Regardless of the above, in the numerical simulations conducted in our previous study using the SPH model, we observed the following phenomena consistent with the phenomenological definition of a vortex lattice: (1) the emergence of multiple vortices spinning in the same direction and (2) their collective rotation around the center of a cylindrical container at a constant speed, i.e., rigid-body rotation. Nevertheless, to prove that the vortex lattice observed in our previous study corresponds to a real-world vortex lattice, it is necessary to theoretically demonstrate the validity of coupling the two motion equations of the two-fluid model, which was experimentally used in our previous studies~\cite{Tsuzuki_2021, doi:10.1063/5.0060605}. In the subsequent sections, we present that the motion equation of an inviscid flow and the nonlinear Schr${\rm \ddot{o}}$dinger equation of a bosonic system in SPH forms are equivalent if certain conditions are satisfied. 
Coupling the motion equation of an inviscid flow and the Navier--Stokes equation is seemingly inadequate to capture the dynamics of quantum fluid phenomena because both these equations are classical equations of motion. 
However, if we prove that the motion equation of an inviscid flow is equivalent to the GP equation in the SPH form under appropriate conditions, we can claim that our coupling method is equivalent to the method that couples the GP equation with the Navier--Stokes equation in such cases. 
Therefore, demonstrating the equivalence of the nonlinear Schr${\rm \ddot{o}}$dinger equation and the motion equation of an inviscid flow in their SPH forms strengthens the validity of coupling the two equations of the two-fluid model from a theoretical perspective.}

\section{Expressions of the GP equation and two-fluid model in SPH forms}
Let us consider a typical case wherein the mutual interaction potential $V$ is given by a scalar multiple of the delta function $\delta$ as follows:
\begin{eqnarray}
V(\vec{r} - \vec{\acute{r}}) &=& g\delta(\vec{r} - \vec{\acute{r}}), \label{eq:UnitConstDef} 
\end{eqnarray}
where $g$ denotes the coupling constant, which represents the interaction strength. 

\HLPOF{Equation (\ref{eq:UnitConstDef}) represents a contact interaction model that assumes weak interactions between atoms. Therefore, it is essentially suitable for reproducing the behavior of dilute gases. Conversely, it is not suitable to reproduce the behavior of superfluid helium-4 based on strong interactions. Therefore, fluid equations derived from the GP equation are often used only to discuss their qualitative behavior. However, in SPH, the delta function is replaced by a kernel function, as in Eq.~(\ref{eq:SPHapprox}). Hence, Eq.~(\ref{eq:UnitConstDef}) acts as a strong interaction model that determines the contribution of particles in the neighboring area with an effective radius. Thus, for SPH calculations, Eq.(\ref{eq:UnitConstDef}) can be expected to provide a relatively accurate approximation of superfluid helium-4 behavior.}

By substituting Eq.~(\ref{eq:UnitConstDef}) into Eq.~(\ref{eq:MomentumEqNThetaiVer}), \HL{which represents} the third term on the right-hand side of Eq.~(\ref{eq:MomentumEqNThetaiVer}) \HL{in a discrete form} using Eq.~(\ref{eq:defDiracPhi}), and then operating $\nabla$ from the left on both sides of Eq.~(\ref{eq:MomentumEqNThetaiVer}), we obtain the following:
\begin{eqnarray}
\frac{\partial \vec{v}_s(\vec{r}, t)}{\partial t} = &-& \frac{\nabla{\vec{v}_{s}(\vec{r}, t)}^2}{2} \nonumber \\
&-& \frac{1}{m_q} \nabla U(\vec{r}, t) \nonumber \\ 
&-& \frac{g}{m_q} \nabla n(\vec{r}, t) \nonumber \\ 
&+& \frac{1}{m_q} \nabla \Bigl[ \frac{\hbar^2}{2 m_q} \frac{\nabla^2 \sqrt{n(\vec{r},t)}}{\sqrt{n(\vec{r}, t)}} \Bigr]. \label{eq:MomentumEqN}
\end{eqnarray}
Here, we use Eq.~(\ref{eq:SupeVelDefPhase}) to derive the left-hand side and first term on the right-hand side of Eq.~(\ref{eq:MomentumEqN}).
The expression provided in the square brackets in the fourth term on the right-hand side is known as the ``quantum pressure,'' hereinafter denoted as $P_q$. 
If the spatial variation in the condensate density profile, $n(\vec{r}, t)$, is small, the quantum pressure can be neglected~\cite{Barenghi2016}. The chemical potential $\mu(\vec{r}, t)$ is then expressed as follows~\HL{\cite{pethick2008bose}}:
\begin{eqnarray}
\mu(\vec{r}, t) &=& U(\vec{r}, t) + g n(\vec{r}, t) - P_q(\vec{r},t), \nonumber \\
P_q &\coloneqq& \Bigl[ \frac{\hbar^2}{2 m_q} \frac{\nabla^2 \sqrt{n(\vec{r},t)}}{\sqrt{n(\vec{r}, t)}} \Bigr]. \label{eq:chemPot}
\end{eqnarray}
The set of second to fourth terms on the right-hand side of Eq.~(\ref{eq:MomentumEqN}) corresponds to $-\nabla \mu/m_q$. Using the relationship $\nabla({\vec{v}_{s}}^2/2) = \vec{v}_s \times (\nabla \times \vec{v}_s) + (\vec{v}_s \cdot \nabla) \vec{v}_s$ and the condition of irrotational flow of the superfluid component $\nabla \times \vec{v}_s = 0$, we can rewrite Eq.~(\ref{eq:MomentumEqN}) as follows:
\begin{eqnarray}
\frac{\partial \vec{v}_s}{\partial t} + (\vec{v}_{s}\cdot\nabla) \vec{v}_{s} = - \frac{\nabla \mu(\vec{r}_{i}, t)}{m_q}. \label{eq:MotionEqChemPot}
\end{eqnarray}
Using Eqs.~(\ref{eq:SPHgraddisc}) and ~(\ref{eq:chemPot}), we can further rewrite Eq.~(\ref{eq:MotionEqChemPot}) in a discretized form of SPH. The resulting equation for the $i$th fluid particle is as follows: 
\begin{eqnarray}
\frac{\partial \vec{v}_s}{\partial t} + (\vec{v}_{s}\cdot\nabla) \vec{v}_{s} = - \frac{\nabla \mu(\vec{r}_{i}, t)^{\rm SPH}}{m_q}, \label{eq:MotionEqChemPotSPH}
\end{eqnarray}
where the right-hand side can be obtained as follows:
\begin{eqnarray}
-\frac{\nabla \mu(\vec{r}_{i}, t)^{\rm SPH}}{m_q} = &-&\sum_{j=1}^{N_p} \frac{U_j}{m_q}\frac{m_j}{\rho_j} \nabla W_{ij} 
\nonumber \\ 
&-& \sum_{j=1}^{N_p} \frac{g n_j}{m_q} \frac{m_j}{\rho_j} \nabla W_{ij} 
\nonumber \\ 
&+& \frac{\nabla P_{q}^{(i)}}{m_q}. \label{eq:MomentumEqNSPHFormFromGP}~~~~~~
\end{eqnarray}
Here, $U_j$ and $n_j$ are the abbreviations of $U(\vec{r}_j,t)$ and $n(\vec{r}_j,t)$, respectively. $P_{q}^{(i)}$ denotes the discretized form of the quantum pressure $P_q$ in Eq.~(\ref{eq:chemPot}), and it can be represented using a series of SPH operators. We denote the quantum pressure gradient force $-\nabla P_{q}^{(i)}/m_q$ as $\vec{f}_q(\vec{r}_i,t)$ for subsequent discussions.
In this manner, by approximating the delta function $\delta$ with the kernel function $W$, we derive the SPH form of the motion equation of condensates from the GP theory. Specifically, this represents the motion equation driven by the gradient of the chemical potential obtained from the Schr${\rm \ddot{o}}$dinger equation of interacting bosons. In summary, we arrange Eq.~(\ref{eq:MomentumEqNSPHFormFromGP}) as follows:
\begin{eqnarray}
\therefore - \left. \frac{\nabla \mu(\vec{r}_{i}, t)^{\rm SPH}}{m_q} \right |_{\rm QM} = &-&\sum_{j=1}^{N_p} \biggr(\frac{gn_j + U_j}{m_q}\biggl)\frac{m_j}{\rho_j} \nabla W_{ij} 
\nonumber \\ 
&-& \vec{f}_q(\vec{r}_i,t). \label{smmary:chempfromGP}
\end{eqnarray}
Here, the subscript QM on the left-hand side symbolically indicates that the right-hand side is an equation derived from the microscopic equation of motion. 

\begin{figure*}[t]
\centerline{\includegraphics[scale=0.52]{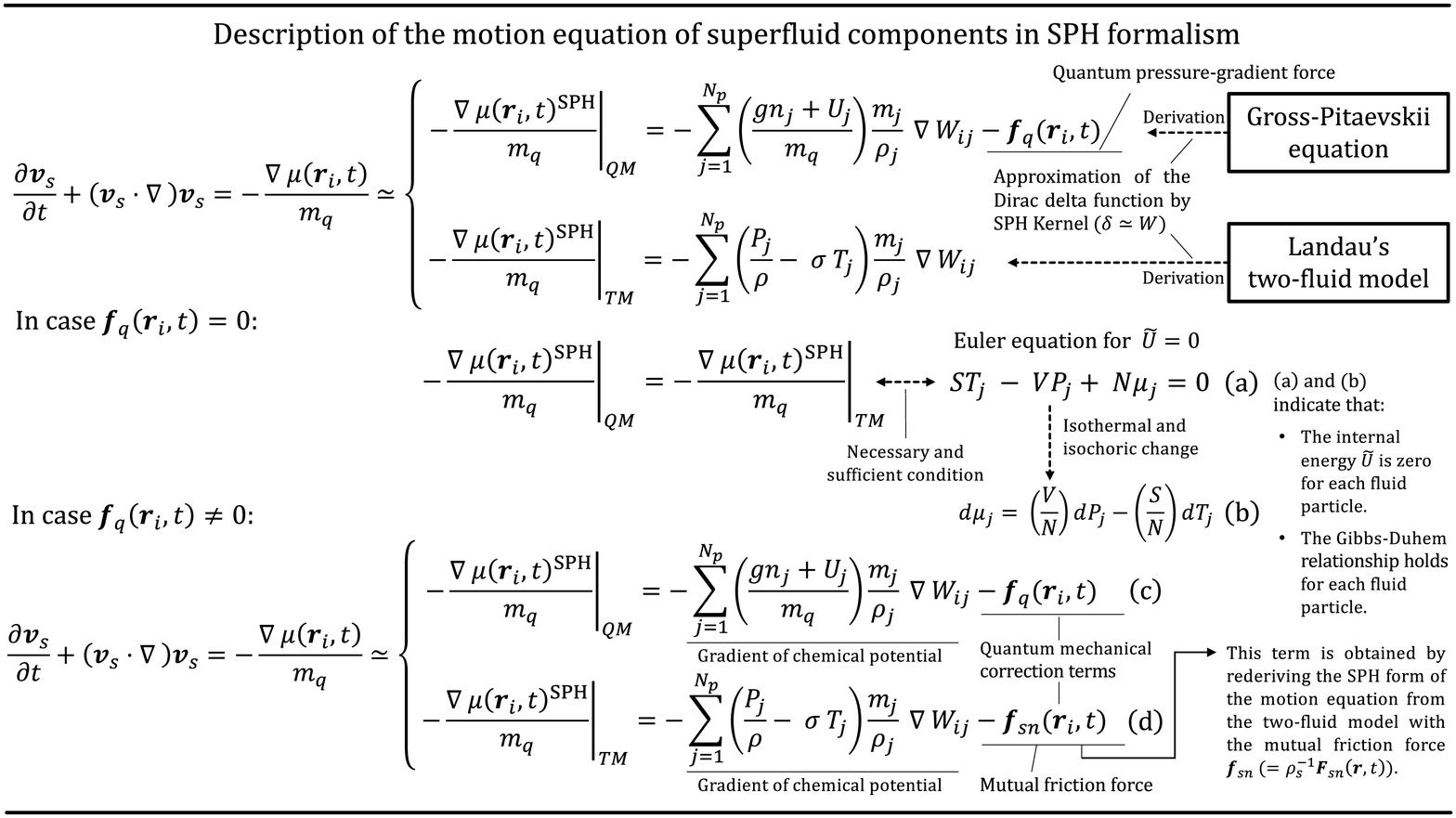}}
\caption{Schematics of the two types of SPH forms of the motion equation of the superfluid component of liquid helium-4 derived from microscopic and macroscopic perspectives.}
\label{fig:OverviewQMandTMSPHModel}
\end{figure*}

Let us recall that the phenomenological equation of motion of the superfluid component of the two-fluid model can be obtained by substituting the Gibbs--Duhem equation, i.e., $\nabla \mu = (V/N)\nabla P - (S/N)\nabla T$, relation $\rho= (Nm_q)/V$ for density, and relation $\sigma = S/(Nm_q)$ for entropy density into Eq.~(\ref{eq:MotionEqChemPot}). In these expressions, $V$ denotes the volume of the entire system, $m_q$ is the mass of a particle (an atom), $N$ is the number of atoms, $S$ is the entropy, and $T$ is the temperature. The resulting equation is as follows: 
\begin{eqnarray}
\frac{\partial \vec{v}_s}{\partial t} + (\vec{v}_{s}\cdot\nabla) \vec{v}_{s} &=& - \frac{\nabla\mu(\vec{r}, t)}{m_q} \nonumber \\ 
&=& - \frac{1}{\rho}\nabla P + \sigma \nabla T. \label{eq:MotionEqTwoFluidSuper}
\end{eqnarray}
Using Eq.~(\ref{eq:SPHgraddisc}), we obtain another SPH form of the gradient of the chemical potential, $\mu$, derived from the Gibbs--Duhem equation, as follows:
\begin{eqnarray}
-\frac{\nabla \mu(\vec{r}_{i}, t)^{\rm SPH}}{m_q} = &-&\sum_{j=1}^{N_p} \frac{P_j}{\rho}\frac{m_j}{\rho_j} \nabla W_{ij} 
\nonumber \\ 
&+& \sum_{j=1}^{N_p} \sigma T_j \frac{m_j}{\rho_j} \nabla W_{ij},
\label{eq:MomentumEqNSPHFormFromTwoFluid}~~~~~~
\end{eqnarray}
where $P_j$ and $T_j$ denote the pressure and temperature of the $j$th fluid particle, respectively. In summary, we obtain the following: 
\begin{eqnarray}
\therefore - \left. \frac{\nabla \mu(\vec{r}_{i}, t)^{\rm SPH}}{m_q} \right |_{\rm TM} = &-&\sum_{j=1}^{N_p} \biggr(\frac{P_j}{\rho} - \sigma T_j\biggl)\frac{m_j}{\rho_j} \nabla W_{ij}. \nonumber \\ \label{smmary:chempfromGD}
\end{eqnarray}
Here, the subscript TM on the left-hand side symbolically indicates that the right-hand side is an equation derived from a thermodynamic viewpoint.
The upper part in Fig.~\ref{fig:OverviewQMandTMSPHModel} presents a schematic of the derivation of the two different SPH forms in Eqs.~(\ref{smmary:chempfromGP}) and ~(\ref{smmary:chempfromGD}).

In the following, we compare the discretized motion equations given in Eqs.~(\ref{smmary:chempfromGP}) and ~(\ref{smmary:chempfromGD}), which are separately derived from a microscopic and macroscopic perspective, to identify a condition under which they become equivalent. Let us focus on a case wherein the spatial variation in the condensate density, $n(\vec{r}, t)$, is sufficiently small; therefore, the quantum pressure, $P_q(\vec{r}, t)$, can be negligible. 
Because $\vec{f}_q(\vec{r}_i,t)$ in Eq.~(\ref{smmary:chempfromGP}) vanishes owing to $P_q(\vec{r},t) = 0$, we can derive a condition for the equivalence of Eqs.~(\ref{smmary:chempfromGP}) and ~(\ref{smmary:chempfromGD}) by comparing each term, as follows:
\begin{eqnarray}
\frac{gn_j + U_j}{m_q} &=& \frac{P_j}{\rho} - \sigma T_j.\label{eq:equivcondorign}
\end{eqnarray}
Using the total density ($\rho = (N m_q)/V$), entropy density ($\sigma = S/(N m_q)$), and chemical potential ($\mu_j = U_j + gn_j+ P_q^{(j)}$) for the $j$th fluid particle, Eq.~(\ref{eq:equivcondorign}) can be represented as follows:
\begin{eqnarray}
\therefore~ST_j - VP_j+ N\mu_j = 0. \label{eq:ResInternalZero}
\end{eqnarray}
Note that $P_q^{(j)}$ of $\mu_j$ is zero in this case. Eq.~(\ref{eq:ResInternalZero}) represents a specific case of the thermodynamic Euler equation~\cite{Basaran2021}, $ST - VP+ N\mu = \tilde{U}$, for the $j$th fluid particle when its internal energy is zero, \HLL{i.e.,} $\tilde{U} = 0$. 
Here, the ~\textasciitilde~\HL symbol {is} added to the internal energy to avoid confusion with the potential energy, $U$. Eq.~(\ref{eq:ResInternalZero}) indicates that the equivalence of Eqs.~(\ref{smmary:chempfromGP}) and ~(\ref{smmary:chempfromGD}) holds when the internal energy of each fluid particle \HLL{is} zero.
In other words, this equivalence holds if the fluid system is in the ground state. Because this is a thermodynamic condition, the ``ground state'' includes an elementary excitation \HLL{state}~(\HLL{refer to} \cite{PhysRev.60.356, Adamenko_2008, bennemann2013novel, schmitt2015introduction} \HLL{for a detailed explanation of the elementary excitation}).
\HLPOF{Herein, the ground state in quantum mechanics is referred to as the ``real ground state.'' We refer to
the ground state derived explicitly from classical thermodynamics as the ``thermodynamic ground state.'' If we deduce that a system is in the thermodynamic ground state based on classical thermodynamics, we cannot conclude that the system is in the real ground state. This is because we cannot deny the possibility that the system is in an elementary excitation state. This information on elementary excitation can only be obtained by considering quantum statistical mechanics. In this regard, Eq.~(\ref{eq:ResInternalZero}) represents a specific case of the thermodynamic Euler equation of each fluid particle when the internal energy is zero. This equation represents the classical thermodynamic relationship. Hence, Eq.~(\ref{eq:ResInternalZero}) suggests that the system is in the thermodynamic ground state. Therefore, there still exists a possibility that the system is in the elementary excitation state. Let us recall that our purpose is to satisfy the relationship in Eq.~(\ref{eq:ResInternalZero}) to establish the equivalence of the two different SPH forms. Particularly, satisfying Eq.~(\ref{eq:ResInternalZero}) indicates that the motion equation of an inviscid flow and the nonlinear Schr${\rm \ddot{o}}$dinger equation of a bosonic system are equivalent in their SPH forms when the density variation is moderate. Once this equivalence is validated, the motion equation of an inviscid flow can be coupled with the Navier--Stokes equation with theoretical validation, unlike in our previous studies, which experimentally adopted the coupling method. As long as the fluid speed does not exceed the critical speed for an elementary excitation, we can assert that the system is in a state close to the real ground state, i.e., a state that can be sufficiently regarded as the ground state from a thermodynamics perspective. This can be achieved when the speeds of all fluid particles do not exceed the critical speed because the total fluid speed is equal to or lower than the speed of an individual fluid particle in SPH. In this case, we can conclude that the system satisfies Eq.~(\ref{eq:ResInternalZero}) for all fluid particles, and hence, the equivalence holds. }

\HLPOF{In addition}, because $V$, $N$, and $S$ are constant parameters, the total differential of Eq.~(\ref{eq:ResInternalZero}) can be obtained as follows: 
\begin{eqnarray}
d\mu_j = (V/N)dP_j - (S/N)dT_j. \label{eq:gibbsduhemforeachp}
\end{eqnarray}
This indicates that the Gibbs--Duhem relationship holds for each fluid particle when Eq.~(\ref{eq:ResInternalZero}) is satisfied. These findings can be summarized as follows: If the spatial variation in the condensate density is sufficiently small such that the quantum pressure is negligible and if the internal energy of each fluid particle remains zero, the motion equation of the superfluid component becomes equivalent to the motion equation of the condensates derived from the GP theory. Here, in the former, fluid particles are driven by the gradient of the chemical potential obtained \HL{using} the Gibbs--Duhem equation, and in the latter, fluid particles are driven by the gradient of the chemical potential obtained from the Schr${\rm \ddot{o}}$dinger equation of interacting bosons. 
The middle part of Fig.~\ref{fig:OverviewQMandTMSPHModel} presents a schematic summary of these results.

If the quantum pressure, $P_q(\vec{r},t)$, is nonnegligible, the quantum pressure gradient force, $\vec{f}_{q}(\vec{r}_{i}, t)$, in Eq.~(\ref{smmary:chempfromGP}) denotes the mutual friction force for the superfluid component of the two-fluid model in a counterflow. This implication is based on the following consideration. 
From a review of previous studies, we can conclude that the velocity distribution in counterflow experiments \HLL{exhibits} a flat profile along the moving direction as the amount of heat input to the system, $W_{in}$, increases~\cite{TOUGH1982133, PhysRevLett.105.045301, Kobayashi2019}. This observation is different from the parabolic profile that follows the Hagen--Poiseuille equation of a laminar flow~\cite{doi:10.1146/annurev.fl.25.010193.000245, doi:10.1063/1.1883163}. 
Specifically, the temperature gradient becomes proportional to $W_{in}$ \HL{cubed} as $W_{in}$ increases. Gorter and Mellink introduced a pair of mutual frictional forces $\vec{F}_{sn}(\vec{r}, t)$ into the two components, as shown in the upper part of Fig.~\ref{fig:OverviewPrevOurWork}, to explain the discrepancy between the results of Landau's two-fluid model and experiments~\cite{GORTER1949285}. Hall and Vinen corroborated that the mutual friction force can be attributed to the interactions between the normal fluid component and quantum vortices~\cite{doi:10.1098/rspa.1956.0214, doi:10.1098/rspa.1956.0215, doi:10.1098/rspa.1957.0071, NEMIROVSKII201385}. In this manner, herein, the mutual friction forces are introduced as quantum mechanical corrections. In our formalism, $-\vec{f}_{sn}(\vec{r}, t) = -(1/\rho_s) \vec{F}_{sn}(\vec{r}, t)$ is added to the right-hand side of Eq.~(\ref{eq:MotionEqTwoFluidSuper}). Hence, the discretized form of the mutual friction force, $-\vec{f}_{sn}(\vec{r}_{i}, t)$, for the $i$th fluid particle is added to the right-hand side of Eq.~(\ref{smmary:chempfromGD}). 

Eqs. (c) and (d) in the lower part of Fig.~\ref{fig:OverviewQMandTMSPHModel} represent Eqs.~(\ref{smmary:chempfromGP}) and ~(\ref{smmary:chempfromGD}), respectively, with the addition of $-\vec{f}_{sn}(\vec{r}_i, t)$. The following important findings are obtained based on these two SPH forms. If we denote the chemical potential at $P_{q}=0$ as $\bar{\mu}$, based on Eq.~(\ref{eq:chemPot}), relation $\mu = \bar{\mu} + P_{q}$ holds, indicating that $\bar{\mu}$ is the classical thermodynamic chemical potential. In addition, from Eq.~(\ref{eq:MomentumEqNSPHFormFromGP}) and relation $\mu = \bar{\mu} + P_{q}$, the first term in Eq. (c) represents the discretized form of the gradient of the classical chemical potential, $\bar{\mu}$, which \HL{naturally} corresponds to the first term in Eq. (d), \HL{owing to the Gibbs--Duhem relation}. In contrast, $\vec{f}_q(\vec{r}_i, t)$ denotes the gradient of the quantum pressure scaled by $-m_q$, and $\vec{f}_{sn}(\vec{r}_i, t)$ denotes the quantum mechanical correction. In brief, the first terms \HL{in} Eqs. (c) and (d) represent the \HLL{discretized} equation of the classical chemical potential, and the second terms \HL{in} Eqs. (c) \HL{and} (d) represent the \HLL{discretized} equation of a quantum mechanical correction given by either $\vec{f}_q(\vec{r}_i, t)$ or $\vec{f}_{sn}(\vec{r}_i, t)$. Therefore, it is reasonable to hypothesize that the equivalence of (c) and (d) is established when each term is equal to the corresponding term. The condition for the equivalence of the first terms in Eqs. (c) and (d) can be derived by comparing each term in the sum, which satisfies the relation $ST_{j} - V P_{j} + N\bar{\mu}_{j} = 0$, where $\bar{\mu}_{j}$ is the classical chemical potential of the $j$th fluid particle. From the Euler equation, we can observe that the equivalence of the first terms holds if the internal energy of each fluid particle is zero. Thus, Eqs. (c) and (d) are equivalent if their second terms are equal. These findings can be summarized as follows. If the quantum pressure is nonnegligible and if the quantum pressure gradient force equals the mutual friction force, the equivalence of the two different SPH motion equations separately derived from a microscopic or macroscopic viewpoint is established. This is valid under the condition that the internal energy of each fluid particle is zero.

\HLPOF{Thus, we determined the condition for the equivalence of the motion equations when the quantum pressure becomes nonnegligible owing to the large spatial variation in the condensate density profile. However, this scenario rarely occurs for most incompressible flows based on SPH because the density fluctuation is typically below 1\% of its averaged density in SPH calculations~\cite{becker2007weakly, MONAGHAN1994399, NOMERITAE2016156}. Therefore, as long as Eq.~(\ref{eq:ResInternalZero}) is satisfied for each fluid particle, the equivalence of the two different SPH forms is almost always guaranteed. Nevertheless, as an exceptional case, the quantum pressure may be nonnegligible at the interface of two different types of fluid particles owing to the differences in their densities. However, the detailed dynamics at such interfaces are yet to be revealed. In the following, we present our perspectives on the equivalence of the two SPH forms at the interface of two different fluid particles.} 

In our opinion, the equivalence of the quantum pressure gradient and the mutual friction forces \HLPOF{at the interface of two different fluid particles} can be maintained, particularly in the case of weakly compressible flows included in explicit SPH simulations of incompressible flows. \HLPOF{A thought experiment on a counterflow can explain the foregoing}. As the heat input increases, the \HLPOF{relative} velocities of both fluid components increase, and the mutual friction forces must be included. Concurrently, in the SPH simulations with an explicit time-integrating scheme, the spatial change in the velocities of fluids generates a density variation in time according to the equation of continuity, \HLPOF{owing to the weak incompressibility}. \HLPOF{This variation is minimal, below 1\% for most parts of the flow, as mentioned above; however, it becomes evident at the interface owing to the density difference and yields a steeper density gradient} with time evolution. Consequently, a quantum pressure gradient force can be induced with the increasing heat input, similar to the case of mutual friction forces. \HLPOF{In this manner, the scenarios in which both forces coexist are similar},and we can \HLPOF{expect} their equivalence to hold to some extent. Interestingly, the weak compressibility in explicit SPH is often regarded as a drawback for classic incompressible fluids; however, it is a significant characteristic that can reproduce the effect of quantum mechanical corrections in quantum fluids. Specifically, the density variation in explicit SPH can create quantum many-particle-interacting systems; this could be one reason for the successful reproduction of the quantum mechanical phenomenon of vortex lattice formation using a fully classical mechanical approximation in our previous study.

\begin{figure*}[t]
\centerline{\includegraphics[scale=0.5]{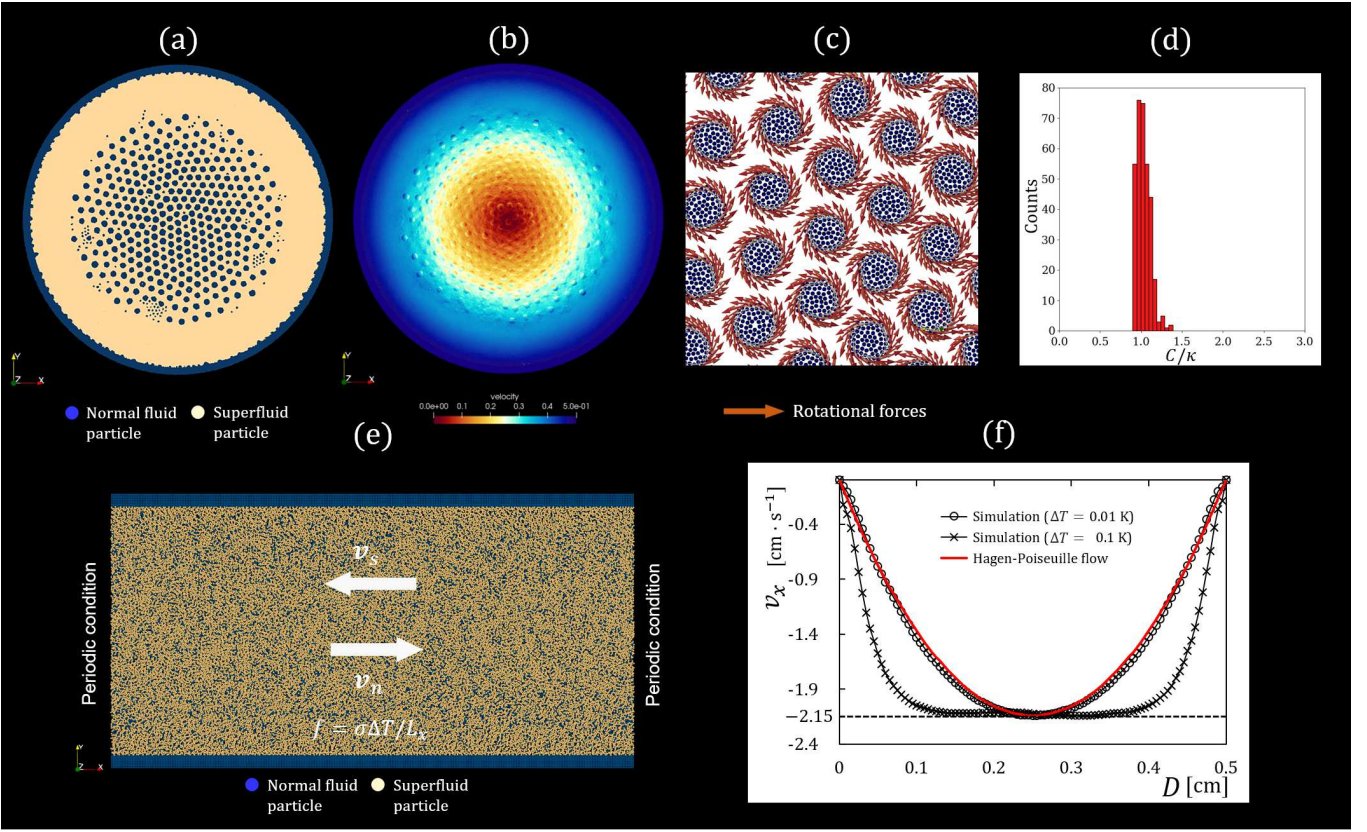}}
\caption{Simulation results of rotating liquid helium-4: (a) two-component view, (b) spatial velocity distribution depicting the local velocity fields of spinning vortices scattered inside a global constant-velocity field, (c) enlarged view of multiple spinning vortices represented in rotational forces, (d) histogram of circulations indicating the quantization of circulations, results for two-dimensional counterflow problems, (e) snapshot of the simulation, and (f) velocity profile in the horizontal direction, which becomes parabolic and flat when $\Delta {\rm t~=~0.01~K}$ and $\Delta {\rm t~=~0.1~K}$, respectively.}
\label{fig:SimulationResults}
\end{figure*}
\section{Application to practical problems} \label{seq:apptopra}
We can assert that the objective of this study has already been accomplished based on the discussion in the previous sections and the following results. If the spatial variation in the condensate density is sufficiently small \HL{such that} the quantum pressure \HL{can be negligible}, the two discretized motion equations in their SPH forms, which are derived separately from the GP equation and the two-fluid model, are equivalent, provided that the internal energy of each fluid particle is zero. \HL{This equivalence holds even when the quantum pressure is nonnegligible if the quantum pressure gradient force equals the mutual friction force.} 
\HLPOF{These results theoretically justify our method of coupling the motion equations of the two-fluid model, which was adopted experimentally in our previous studies. Coupling the motion equation of an inviscid flow and the Navier--Stokes equation is seemingly inadequate to capture the dynamics of quantum fluid phenomena because both these equations are classical equations of motion. However, at least in the SPH form, because the motion equation of an inviscid flow is equivalent to the GP equation under appropriate conditions, our coupling method is equivalent to the method that couples the GP equation with the Navier--Stokes equation in such cases. The latter method is seemingly more consistent with well-accepted methods that couple the quantum mechanical equations with the Navier--Stokes equation than the former, which couples the motion equation of an inviscid flow with the Navier--Stokes equation. However, our analysis indicates that they are equivalent, at least in their SPH forms, as long as the internal energy of each fluid particle remains zero when the quantum pressure is negligible. In summary, our method of coupling the motion equation of an inviscid flow with the Navier--Stokes equation is valid for simulating liquid helium-4 if the aforementioned condition is satisfied. Therefore, in this section, we present the use of this method in simulating the dynamics of liquid helium-4, as in our previous studies.}
As a developmental step in our study, we discuss the implications of these findings for solving real-world cases and their incorporation into liquid helium-4 simulations using our two-fluid model to accurately reproduce actual phenomena.

\HL{As previously mentioned}, the condition requiring that the internal energy of a fluid particle be zero \HLL{implies} that the fluid particle is in the thermodynamic ground state. Because this is a thermodynamic condition, the ground state includes an elementary excitation state. One approach to sufficiently satisfy this condition is to ensure that the velocity of each fluid particle does not exceed the Landau critical velocity. Under the assumption of a weakly interacting Bose--Einstein condensate, it is generally known that the magnitude of the Landau critical velocity is given by the speed of sound~\cite{PhysRevB.105.014515}. For helium-4, the critical velocity becomes lower than the speed of sound owing to the presence of rotons~\cite{donnelly1991quantized, PhysRevB.103.104516}. Although this depends on the problem and measurement method, experiments involving highly sensitive flow measurement systems based on diaphragm displacement with a DC detection circuit have revealed that the critical \HL{velocity} generally ranges from 7.7 $\rm m\cdot s^{-1}$ to 11.6 $\rm m\cdot s^{-1}$. Moreover, it ranges from approximately 0.16 $\rm m\cdot s^{-1}$ to 0.22 $\rm m\cdot s^{-1}$ in the slowest case observed during certain irregular transition states~\cite{lindensmith1996critical}. Let us now focus on flow problems wherein the characteristic velocity of the flow is given by a few $\rm cm\cdot s^{-1}$, as a typical case of laboratory systems. Here, the fluid speed is always lower than the critical velocity when considering the simulations of liquid helium-4 at the laboratory scale. \HL{Therefore}, \HL{the condition for equivalence can naturally be satisfied.}	

In addition, in quantum hydrodynamics, the circulation, $C$, of a vortex is given by an integer multiple of the quantum of circulation, $\kappa$, as expressed by $C=q\kappa$. Such a topological defect enables vortices to maintain dynamic stability. Here, $q$ is typically one, and a situation wherein the circulation becomes more than twice $\kappa$ is rarely encountered because the higher energy states are unstable. In the rotation problem, a vortex lattice is formed owing to the balance between the repulsive forces acting among multiple parallel vortices and the Magnus force. The latter acts as a centripetal force owing to the forced rotation of the outer vessel. 
Accordingly, in the simulations of rotating liquid helium-4, it is necessary to ensure that the magnitude of the circulation in each vortex is approximately equal to $\kappa$ to accurately reproduce the effect of quantization of circulation. In our previous study~\cite{doi:10.1063/5.0060605}, we \HL{monitored} the circulation of clusters formed by accumulation of fluid particles at every time step. Therein, if the circulation of a cluster was greater than the quantum of circulation, $\kappa$, it was identified as a vortex. In this case, we obtained the repulsive forces based on the vortex dynamics \HL{theory}~\cite{doi:10.1063/5.0060605, pethick2008bose} between the vortex and other vortices or clusters to prevent further vortex growth. Consequently, we succeeded in reproducing the phenomenon of vortex lattice formation; however, we obtained a histogram wherein the circulation of each vortex was continuously distributed for circulation values greater than 1--3 times the $\kappa$ value. However, we failed to quantize the circulation; therefore, we could not regard the formed vortex lattice as a ``quantum lattice.'' As a remedial measure, in this study, we apply a stronger constraint on the system; under this constraint, repulsive forces are applied to the interactions among clusters in advance to prevent them from merging when the sum of the circulations of two approaching clusters is estimated to be greater than $\kappa$ to reproduce topological defects, and the constraint is expressed as follows:
\begin{eqnarray}
 F^{(k,l)}_{int} = 
~\Biggl\{ \begin{array}{l}
{0~~~~~~\Biggl( \begin{array}{l} \Gamma_{k} < w_{c}~q \kappa, \\ \Gamma_{l} < w_{c}~q \kappa, \\ \Gamma_{k} + \Gamma_{l} < \gamma_{c}~q \kappa \end{array}\Biggr) } \\
{f^{(k,l)}_{int}~~{\rm (otherwise),}} 
	\end{array}
	\label{eq:vvforcejudgementmodif}
\end{eqnarray}
where $ F^{(k, l)}_{int}$ denotes the resulting interaction force between the $k$th and $l$th clusters or vortices, and $f^{(k,l)}_{int}$ denotes the repulsive force of the vortex dynamics model. We use the same $f^{(k,l)}_{int}$ as in $\mathrm{I}\hspace{-1.2pt}\mathrm{I} \hspace{-1.2pt}\mathrm{I}.~{\rm D}$ in~\cite{doi:10.1063/5.0060605}. $\Gamma_{x}~(x=k,l)$ represents the circulation of the $x$th cluster or the vortex. $w_{c}$ and $\gamma_{c}$ are model parameters that confine the circulation of a vortex at approximately the quantum of circulation, $\kappa$, and they are set to approximately 0.9 and 1.1, respectively. \HL{Eq. (\ref{eq:vvforcejudgementmodif}) can prevent vortices from accelerating and exceeding the critical velocity because of the unbalanced interactions among them, which result from the emergence of unrealistic vortices with circulations much greater than $\kappa$.} 

We performed a numerical simulation of rotating liquid helium-4 using SPH under similar computational conditions as in~\cite{doi:10.1063/5.0060605}, except for the conditional judgment represented in Eq.~(\ref{eq:vvforcejudgementmodif}). Normal and superfluid particles were randomly distributed according to their density ratios of $\rho_s/\rho$ and $\rho_n/\rho$ in a cylindrical container with an outer diameter of 0.2~cm; this container began to rotate at a speed of ${\rm 5~rad~s^{-1}}$ after the simulation was initiated. The temperature was maintained at $\rm 1.6~K$ throughout the simulations. Note that practical simulations using SPH require the introduction of several established techniques to ensure numerical stability. Additional details on these techniques can be found elsewhere $\mathrm{I}\hspace{-1.2pt}\mathrm{I}.{\rm ~C}$ in \cite{doi:10.1063/5.0060605}. 
\HLPOF{For the boundary conditions, we first considered the concept of the conventional two-fluid model in the target system and then approximately reproduced the mechanics of our model. Specifically, we assumed a situation wherein the normal fluid component of the conventional two-fluid model is arranged alongside the outer vessel. In this scenario, the wall can interact with the normal fluid component via the viscosity force, and it can also interact with the superfluid component via the mutual friction force. Therefore, in general terms, the walls are in nonslip states against the fluid component. Based on this consideration, we applied a nonslip boundary condition on the system by arranging viscid particles alongside the outer vessel. When the rotation of the outer vessel began, we imposed the velocity of the vessel as a moving boundary condition to the wall particles.}
The results are presented in the upper \HLPOF{row} of Fig.~\ref{fig:SimulationResults}(a)--(c). These illustrate the two-component view, the spatial velocity distribution depicting the local velocity fields of the spinning vortices scattered inside a global constant-velocity field, and an enlarged view of multiple spinning vortices represented by rotational forces, respectively. Similar to ~\cite{doi:10.1063/5.0060605}, it is confirmed that multiple spinning vortices form a rigid-body lattice rotating under a constant-velocity field comprising the local velocity fields of spinning vortices that are indifferent to the motion of the overall velocity field. 
\HLPOF{In more detail, similar to our previous study, we observed small clusters close to the periphery of the vortex lattice. These clusters are bundles of fluid particles that do not grow into vortices; we cannot regard them as vortices because they are just clusters that do not spin. The physical significance of these clusters before they develop into vortices is still under discussion. However, we also observed small clusters of normal fluid particles that were initially attached to the wall and began to detach from the wall owing to the centripetal force resulting from the forced rotation of the outer cylinder.}

Furthermore, we obtained a histogram of circulations, which is presented in Fig.~\ref{fig:SimulationResults}(d), where the horizontal axis indicates the circulation scaled by the quantum of circulation, $\kappa$, and the vertical axis indicates the number of vortices. This figure reveals that all circulations of the vortices are distributed within a range of ${\rm 0.9~\kappa}$--${\rm 1.4 \kappa}$, which is slightly wider but almost similar to the experimental result~\cite{vinen1961detection}. In this study, the circulations appear to be quantized at approximately $\kappa$. 
\HLPOF{We note the existence of a small amount of measured data in the region where $C/\kappa$ is smaller than one in the experimental results, which is not reproduced in the present simulations. Notably, an accurate reproduction of the histogram of circulation depends on Eq.~(\ref{eq:vvforcejudgementmodif}). Our previous study~\cite{Tsuzuki_2021} has already revealed that the simple SPH discretization of the equations of the two-fluid model with angular momentum conservation can reproduce the rigid-body rotation of multiple vortices spinning in the same direction, even without using a vortex dynamics model. In a subsequent study~\cite{doi:10.1063/5.0060605}, we formulated the vortex dynamics in SPH, describing the Magnus force generated by the rotation of each vortex and the repulsive force resulting from the interaction between vortices spinning in the same direction. The problem was to identify a basis for the determination of conversion of \HLPOF{a gathering of} particles into a vortex. Because vortex circulation is generally quantized to approximately $q=\kappa$, we used the aforementioned vortex dynamics model when the circulation \HLPOF{of a gathering of particles} was $q=\kappa$. Consequently, we generated a vortex lattice in our previous study~\cite{doi:10.1063/5.0060605}. However, the quantization of circulation failed because numerous vortices with larger circulations than those observed in the experiments were generated. In a continuous model like the SPH, even if the vortex dynamics forces are applied to clustering particles after the system recognizes them as a vortex, the clustering process proceeds before the applied repulsive force effectively acts on them. Based on this, Eq.~(\ref{eq:vvforcejudgementmodif}) improves the conditional judgment in~\cite{doi:10.1063/5.0060605} to be more sensitive to changes in the circulations of vortices. We can state that the conditional judgment in Eq.~(\ref{eq:vvforcejudgementmodif}) is still tentative and needs further sophistication, primarily from theoretical aspects. Nevertheless, it is still significant that we succeeded in quantizing the circulations of vortices based on our continuum mechanical approach.}
\HLPOF{We also note that the experimental values are for 1.3 K. At this stage of our study, the aim is to qualitatively compare the behavior at similar low temperatures. The fact that the circulation is quantized at approximately $C/\kappa=1$ still remains valid, and we believe that this degree of temperature difference does not affect this discussion within the scope of this study. Nevertheless, a more quantitative comparison based on experiments at the same temperature is planned for the future.}

\HLPOF{We previously demonstrated that in our system, by adjusting the parameter controlling the magnitude of the angular velocity around the axis of each fluid particle ($C_{w}$ in~\cite{doi:10.1063/5.0060605}), the number of generated vortices could match the theoretical value estimated by Feynman's rule. However, the sizes of the vortex lattice and its vortices were different from the experimental values~\cite{PhysRevB.20.1886, PhysRevB.90.174512}. Our current view on this point can be detailed as follows. Let us recall that the balance between the Magnus force generated by the spinning of the vortices and the repulsive force between the vortices determines the vortex arrangement. As the angular velocity of fluid particles become small, the magnitudes of the Magnus force, repulsive force, and circulations of the fluid particles forming each vortex reduce because they depend on the angular velocity, and hence, the system forms a smaller vortex lattice. At this time, if the angular velocity of each fluid particle becomes lower than a certain critical angular velocity, the rotational energy of the vortex becomes insufficient, and the vortex is unable to spin; consequently, it is no longer a vortex. By contrast, if the fluid particles are large, i.e., if the resolution of the system is insufficient, the rotational energy of each vortex in the calculation is larger than that of a real vortex. Nevertheless, we can maintain sufficient rotational energy to observe vortex spinning by setting a higher angular velocity for each fluid particle. In summary, the scale of a vortex lattice can be controlled by the radius of a fluid particle, i.e., the resolution applied to the system, and the magnitude of the angular velocity of a fluid particle. This suggests that our simulation results display a vortex lattice magnified by the amplified angular velocities of the fluid particles. In all cases, it is necessary to ensure that the velocity of a fluid particle does not exceed the critical speed. To ensure the foregoing, the angular velocity considered is the angular velocity of classical fluid particles, with no quantum mechanical significance. Their interpretation in terms of quantum mechanics should be defined in future studies.}

\HLPOF{In these numerical experiments, the density ratio of the two different virtual fluid particles causes the low-density normal fluid particles to aggregate and form vortices. As explained in Section~\ref{sec:explainsph}, all fluid particles considered in our calculation are classical fluid particles that only act as discretization points of the helium-4 continuum. The two types of virtual particles (superfluid and inviscid) have no physical significance other than serving as discretization points, or fragments, of a continuous two-phase flow. Particularly, they are virtual particles. This study adopted a one-fluid system, wherein we focus on the fluid phenomena of the system as a whole; the motions of individual virtual particles have no physical significance from a microscopic viewpoint. It is possible to display these two types of virtual fluid particles separately; nevertheless, our focus is only on the entire fluid field. Therefore, a vortex cannot be decomposed into physically significant quantum components in our current model. However, in future studies, we can possibly represent the internal structure of a vortex by introducing subscale particles inside it and by modeling their interactions with microscopic Lagrangian dynamics, such as molecular dynamics, or by combining with the DFT. In particular, the SPH formulation is well suited to be combined with other Lagrangian approaches for this multiscale physics.}

Furthermore, we simulated two-dimensional counterflow problems. We focused on a simplified scenario wherein a periodic condition was applied to a rectangular domain. Specifically, we set the simulation domain of ($L_x$, $L_y$) as (1~cm, 0.5~cm) and set periodic boundary conditions along the x direction. The characteristic velocity, \HLPOF{which is the maximum velocity of the fluid particles calculated by multiplying the Mach number with the speed of sound, was} ${\rm 2.4~cm \cdot s^{-1}}$. We set the average temperature as ${\rm 1.6~K}$ and the temperature difference, $\Delta t$, as ${\rm 0.01~K}$ or ${\rm 0.1~K}$ in the x direction. Fluid particles were driven by a constant-temperature gradient calculated from the input temperature difference. The other input parameters were the same as those in the rotation problem. To satisfy Landau's criterion, we measured the velocity profile along the x direction when the velocities of the fluid particles were lower than the characteristic velocity. 
Fig.~\ref{fig:SimulationResults}(e) presents \HL{a snapshot of the simulation}. As shown in Fig.~\ref{fig:SimulationResults}(f), the velocity profile in the horizontal direction is confirmed to be parabolic or flat when $\Delta t=0.01~K$ or $\rm \Delta t = 0.1~K$, respectively. 
\HLPOF{The velocity profile appears as a center-flattened profile in the x direction when the amount of heat input to the system increases. This profile characteristic is qualitatively consistent with the results of experiments~\cite{PhysRevLett.105.045301} and numerical simulations~\cite{Saluto2014, PhysRevLett.120.155301} of the normal fluid component often observed in a deformation state~\cite{PhysRevLett.120.155301}. Meanwhile, a tail-flattened profile is observed in the transition regime from the laminar to turbulent flow~\cite{PhysRevB.91.094503}. Thus, our SPH model can be instrumental in exploring the detailed mechanism and conditions under which these different profiles are observed. In particular, reproducing a tail-flattened profile is one future task; however, improving the boundary conditions may be necessary in such cases.}
\HLPOF{A straightforward explanation for the simulation results of the counterflow can be provided from the viewpoint of classical fluid dynamics as follows}. In a fully classical approximation including the fluid forces of both components and their interactions, the normal fluid component acts as a drag force against the superfluid component. In this test case, when we increased the temperature difference from $\Delta t=0.01~K$ to $\rm \Delta t = 0.1~K$, the drag force increased and prevented the entire fluid from growing to create a parabolic profile before the fluid reached the critical \HL{velocity}. In summary, our numerical scheme based on SPH was demonstrated by two representative flow problems: rotation and counterflow problems of cryogenic liquid helium-4. 

The resolutions of each simulation were approximately 200,000 particles for the rotation problem and \HL{52,400} particles for the counterflow problem. All simulations were performed on a GPU NVIDIA Geforce RTX 2080 Ti system. However, it is likely that larger numerical simulations with higher resolutions may capture more precise phenomena. \HLPOF{In particular, in this study, we did not examine the internal dynamics within each vortex. This was because the stability of vortices had to be discussed after the development of a theoretical scheme for the inner structure of vortices in our SPH model. Nevertheless, these detailed dynamics should be explored in numerical simulations with a sufficiently high resolution to capture the internal dynamics of a vortex. Another approach could be to combine our approach with other methods, e.g., the DFT, which is suited to describe the underlying physics in such a subscale regime, instead of simply applying SPH to the two-fluid model. As mentioned in ~\ref{sec:explainsph}, we can interpret fluid particles in SPH not necessarily as physically meaningful particles but as analytic points. Based on this, the differential operators in SPH can also be used in a DFT model. Nevertheless, it still appears promising to discretize the two-fluid model, conserving the angular momentum with ultra-scale resolutions using billions of fine-grained analytic particles to capture the dynamics of vortex nucleation or the inner structure of vortices. 
By comparing the vortex profiles from the high-resolution SPH with those from the DFT calculations, we can also expect to obtain the quantum physical meaning and optimal values of the model parameters of the SPH model, such as $C_w$, which controls the magnitude of the angular velocity around the axis of each fluid particle.
We acknowledge that a few billion analytical particles can be used in SPH calculations within a reasonable computational time on recent supercomputers~\cite{Tsuzuki:2016:EDL:3019094.3019095, 0d91db780fed40d580d4ea4b9e261b7b}; thus, we can state that SPH has a significant potential for diverse approaches. In addition, three-dimensional numerical simulations are required to explore the dynamics of vortex reconnection~\cite{doi:10.1063/1.4772198} or vortex rings~\cite{doi:10.1063/1.5047471}, or the dynamics of the rotating droplets in a spherical configuration~\cite{PhysRevLett.124.215301}. To achieve this, it is essential to develop multi-GPU simulation codes that can efficiently realize computationally expensive large-scale simulations involving billions of particles; however, it is necessary to approach this problem from all angles in science and engineering.}

\section{Conclusion}
Apart from our own previous study, no other study has attempted to reproduce the dynamics associated with macroscopic quantum fluid phenomena using a fully classical mechanical approximation model. Our study was motivated by a purely academic interest aimed at directly reproducing macroscopic quantum phenomena, such as a film flow and the fountain effects.
Our previous studies revealed that a vortex lattice in rotating liquid helium-4 can be reproduced even when the two-fluid model is solved in a fully classical mechanical approximation that includes the fluid forces of both components. This is valid under the condition that the viscosity is rederived to conserve the rotational angular momentum using SPH and the vortex dynamics \HLPOF{is incorporated into the system}. Furthermore, a fully classical mechanical approximation of the two-fluid model using SPH may be equivalent to solving a many-body quantum mechanical equation under specific conditions. However, no study has provided theoretical evidence supporting the existence of this equivalence of the microscopic motion equation of a quantum many-body system and the phenomenological motion equation of the superfluid component of the two-fluid model in an SPH formalism.

This study demonstrated the existence of this equivalence in the following manner. We first derived the SPH form of the motion equation of the superfluid component of the two-fluid model, i.e., the motion equation driven by the gradient of the chemical potential obtained using the Gibbs--Duhem equation. Further, we derived the SPH form of the motion equation of condensates from the GP theory, i.e., the motion equation driven by the gradient of the chemical potential obtained from the Schr${\rm \ddot{o}}$dinger equation of interacting bosons. We then compared these two discretized motion equations in their SPH forms, which were separately derived from a microscopic or macroscopic perspective, to identify a condition of their equivalence. Consequently, we discovered that the thermodynamic condition requiring the internal energy of each fluid particle to be zero ensures this equivalence when the quantum pressure is negligible. Our results also indicated that this equivalence holds even when the quantum pressure is nonnegligible if the quantum pressure gradient force equals the mutual friction force. 

\HLPOF{These results theoretically justify our method of coupling the motion equations of the two-fluid model, which was adopted experimentally in our previous studies. Coupling the motion equation of an inviscid flow with the Navier--Stokes equation is seemingly inadequate to capture the dynamics of quantum fluid phenomena because both these equations are classical motion equations. However, at least in their SPH forms, because the motion equation of an inviscid flow is equivalent to the GP equation under appropriate conditions, our coupling method is equivalent to the method of coupling the GP equation with the Navier--Stokes equation in such cases. The latter method is seemingly more consistent with well-accepted methods that couple quantum mechanical equations to the Navier--Stokes equation than the former, which couples the motion equation of an inviscid flow with the Navier--Stokes equations. However, our analysis indicated that they are equivalent, at least in their SPH forms, as long as the internal energy of each fluid particle remained zero when the quantum pressure was negligible. In summary, our method of coupling the motion equation of an inviscid flow with the Navier--Stokes equation was demonstrated to be valid for simulating liquid helium-4 under a scenario wherein the aforementioned condition was satisfied. Therefore, it is still acceptable to adopt the method of coupling the motion equation of an inviscid flow with the Navier--Stokes equation in simulating the dynamics of liquid helium-4, as in our previous study.}

\HLPOF{In addition to the case of negligible quantum pressure, we also discussed the condition for the equivalence of the equations when the quantum pressure was nonnegligible owing to the large spatial variation in the condensate density profile. However, this scenario is rarely encountered for most incompressible flows based on SPH because the density fluctuation is typically maintained below 1\% compared to the averaged density in SPH. Thus, as long as the internal energy of each fluid particle is zero, the equivalence of the two differential SPH forms is almost always guaranteed. However, in exceptional cases, the quantum pressure may be nonnegligible at the interface of two different types of fluid particles owing to the difference in their densities. However, the detailed dynamics at such an interface are still under discussion and require further study.}

The condition requiring each fluid particle to have a zero internal energy can naturally be satisfied in several numerical simulations with a characteristic velocity of a few $\rm cm\cdot s^{-1}$ for a laboratory system. This condition indicates that each fluid particle is in the thermodynamic ground state. Because this is a thermodynamic condition, the ground state includes an elementary excitation state. One approach to sufficiently satisfy this condition is to ensure that the velocity of each fluid particle does not exceed the Landau critical velocity. For liquid helium-4, several experiments have indicated that the critical \HL{velocity} generally ranges from 7.7 $\rm m\cdot s^{-1}$ to 11.6 $\rm m\cdot s^{-1}$. Moreover, it ranges from approximately 0.16 $\rm m\cdot s^{-1}$ to 0.22 $\rm m\cdot s^{-1}$ in the slowest case. Let us focus on flow problems wherein the characteristic velocity of the flow is given by a few $\rm cm\cdot s^{-1}$, as a typical case of laboratory systems. Here, when considering the simulations of liquid helium-4 at the laboratory scale, the fluid speed is always lower than the critical velocity. Therefore, we performed a simulation of rotating liquid helium-4 with a sophisticated constraint such that the velocities of the fluid \HL{particles} did not exceed the Landau critical velocity. Consequently, we generated a vortex lattice with quantization of the circulation, known as a quantum lattice.

Our method provides a perspective on the dynamics of liquid helium-4 based on a fully classical mechanical approximation that is different from the descriptions obtained using conventional methods. Our theory and simulation results demonstrate that our proposed scheme can be used to appropriately describe the macroscopic dynamics of liquid helium-4. Particularly, a vortex lattice with quantized circulation, i.e., a quantum lattice, can be reproduced by this method based on classical fluid mechanics, although this has been previously considered as a purely quantum mechanical phenomenon. 
We expect that our approach provides a new methodology for describing cryogenic liquid helium-4.







\section*{Acknowledgment}
This study was supported by JSPS KAKENHI Grant Number 22K14177. 
The author would like to thank Editage (www.editage.jp) for English language editing.
The author specially acknowledges Prof. Katsuhiro Nishinari and the administrative staff at the Nishinari Laboratory and the RCAST, University of Tokyo.
The author is also grateful to his family for their warm encouragement.

\bibliographystyle{h-physrev3}
\bibliography{reference}

\end{document}